\documentclass{ws-p8-50x6-00}

\begin{document}

\title{MAXIMUM NON-PERTURBATIVE STRONG INTERACTION HYPOTHESIS AND  THE STRUCTURES OF POMERON AND GLUEBALL }

\author{HONG-AN PENG}

\address{Department of Physics, Peking University, 
 Beijing 100871, China }


\maketitle

\abstracts{In this paper we first emphasize the importance of the Pomeron (IP) for its asymptotically  saturation the unitarity condition alone. After proper modified the field theory model for IP developed by Landshoff and Nochtmann, we argue that the exchange of IP in high energy $h-h'$ scattering embodies the hypothesis of the maximum non-perturbative strong interaction reaction (MNSIR) in which a constituent quark converts into a current quark and emits a color octet non-perturbative gluer. We think the IP is composed from the conjugated pair of such gluer. In the circle of non-perturbative fundamental entities in QCD, we argue that there should exist a new member --- the constituent gluon, it should be emerged in high energy $h-h'$ strong-soft processes, and the glueballs are produced via two constituent gluons fusion. Finally, we conjecture there may be an averaged dual relation between the glueballs and the Pomeron, but this correspondence may become as multi-to-one homologue.}

The Pomeron (IP) plays the most important role in high energy strong-soft processes, 
but it is also yet a poor understanding object in QCD.

Let us observe the diagrammatic equations of unitarity condition for 
$\sigma^T_{h-h'}(s)$ and its Regge formalism.See Fig.a.

\begin{figure}[t]
%\figurebox{20pc}{15pc}{} % to have a box alone
\begin{center}
\epsfxsize=30pc 
\epsfbox{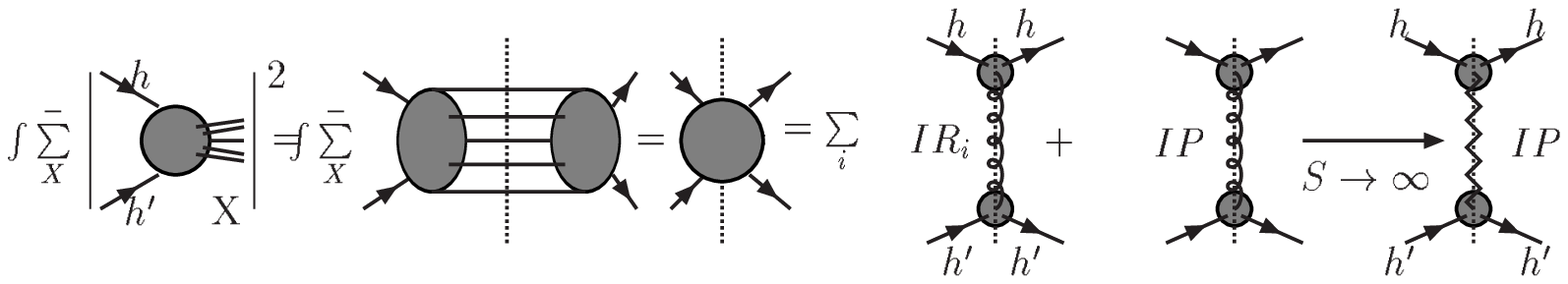}

Fig.a
\end{center}
%\caption{A generalized cactus tree: the confluent
%transfer-matrix $S$ transforms the state function $f(x)$ and
%$f(z)$ into $f(x)$.  \label{fig:radish}}
\end{figure}

The IP term saturated asympototically the unitarity condition alone! It means the 
IP exchange covered all permitted strong-interactions, none other object could compete with it.

The structure of IP hv been studied by Landshoff and Nachtmann (L-N). They take notice 
the additive quark rule from high energy $\pi-N$ and $N-N(\bar N)$ scattering data.
$$\frac{\sigma_{\pi^{-}-P}^T (S)}{\sigma_{\bar P-P}^T(S)}\approx
\frac{\sigma_{\pi^{-}-P}^T (S)}{\sigma_{P-P}^T(S)}\approx
\frac{\sigma_{\pi^{+}-P}^T (S)}{\sigma_{\bar P-P}^T(S)}\approx
\frac{\sigma_{\pi^{+}-P}^T (S)}{\sigma_{P-P}^T(S)}\approx \frac{13.63}{21.70}
\approx \frac{2}{3}$$
This reveals the IP couples only with the constituent quark ($q$ or $\bar q$) in $\pi$
and $N$ , it also means at each one time only one constituent quark in hardron are 
coupled with IP. See Fig.b.

\begin{figure}[t]
%\figurebox{20pc}{15pc}{} % to have a box alone
\begin{center}
\epsfxsize=25pc 
\epsfbox{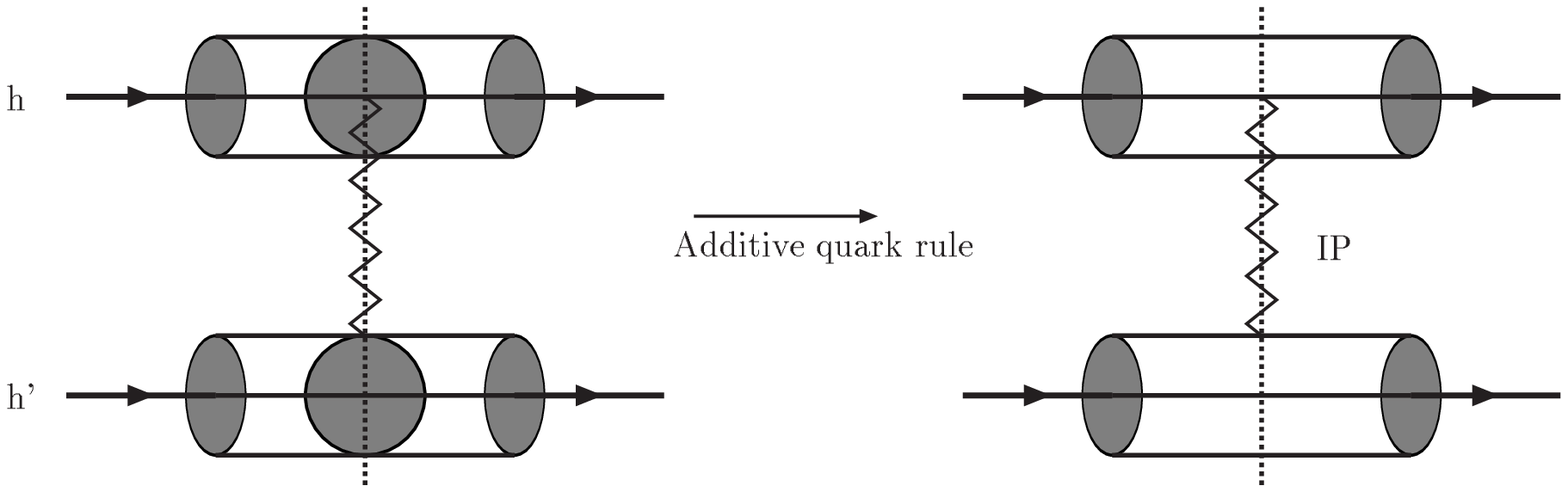}

Fig.b
\end{center}
%\caption{A generalized cactus tree: the confluent
%transfer-matrix $S$ transforms the state function $f(x)$ and
%$f(z)$ into $f(x)$.  \label{fig:radish}}
\end{figure}

L-N IP-model take an Abelian gluon-quark theory, "graft" the relevant non-Abelian 
features of QCD as extra conditions for Abelian cases.

$L-N$ have demonstrated that the IP exchange corresponds two non-pertubative gluons 
exchange and this two-gluon coupled predominately to the same quark in the hardron  in Fig.1

\begin{figure}[t]
\begin{minipage}[t]{.45\linewidth}
\begin{center}
%\vskip -5mm
\epsfxsize=12pc
\epsfbox{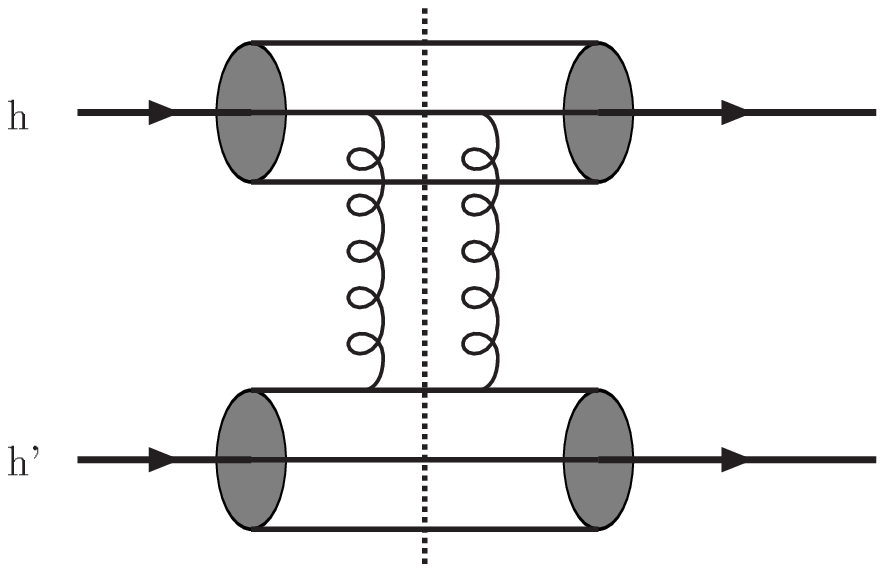} 
\end{center}
%\vskip -5mm
\caption{ Imaginary part of IP exchange in L-N model.
\label{f1}}
\end{minipage}%
\hspace{0.5in}
\begin{minipage}[t]{.45\linewidth}
\begin{center}
%\vskip -5mm
\epsfxsize=12pc
\epsfbox{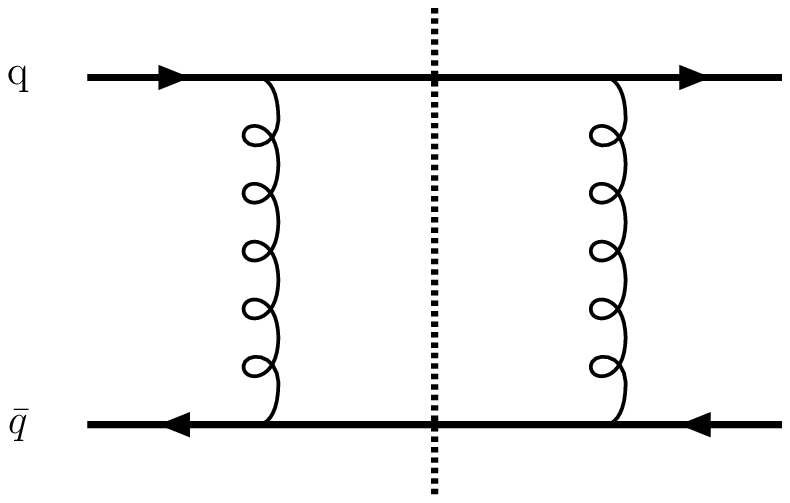}
\end{center}
%\vskip -5mm
\caption{ Sub-diagrams for constituent quark $q-\bar q$ scattering from Fig.1.
\label{f2}}
\end{minipage}
\end{figure}
%\vskip -5mm

The core part of L-N model is Fig.2 

The field theory of IP developed by $L-N$ has well explained the properties of IP 
observed in experiments and has much inspired us. But it is hardly both to excavate 
exhaustingly the physical connotation for IP-exchange and to understand why it  could 
be asymptotically saturated alone the unitarity condition if we just copy whole the 
L-N model. So we propose to modify this model properly.

But how? Pause here and consider the nucleon structure. There are two pictures:

\begin{enumerate}
\item Constituent quark $q$ picture be suited for static or low-$Q^2$ (small $|t|$).
\item Current quark $q_c$ (parton distribution) picture in PQCD be suited for high-$Q^2$.

\end{enumerate}

Could be "link-up" these two picture? It seems very difficult in any DIS
process. If there were any way to realize such "link reaction", perhaps it would be happened 
in high energy strong-soft processes.

Especially we believe the $\sigma^T_{h-h'}(s)$ production processes are "nice place" 
to realize such $q-q_c$ link reaction! Why? Just in it the IP exchange alone 
saturated the unitarity condition. According to constituent quark models in low-$Q^2$  hadron physics,
and QCD arguments, we conjecture the constituent quark($q$) in a hadron is dressed 
with clouds which are composed predominately from non-pertubertive gluons, while as  
the current quark ($q_c$) does not.

Under suitable environment, the non-pertubertive clouded-dress(c) of a constituent 
quark my be stripped out partially or even wholly. If the clouded dress is wholly 
stripped out , i.e. $q\rightarrow q_c+c$, we call this reaction as maximum 
non-perturbative strong interaction reaction (MNSIR). Here $q_c$ is  a "bare" constituent quark but which is just a physical quark in PQCD.

The $\sigma^T_{h-h'}(s)$ production process is just a suitable environment for MNSIR, 
since the $\sigma^T_{h-h'}(s)$ comes predominately from diffractive-like scattering 
distributions which are typically non-pertubative process and surely involved 
largest permitted apace-time($\sim$ 1fm) scale.

We suggest Fig.2 should be modified as Fig.3, its meaning is clear.

\begin{figure}[htb]
%\figurebox{20pc}{15pc}{} % to have a box alone
\begin{center}
\epsfxsize=15pc 
\epsfbox{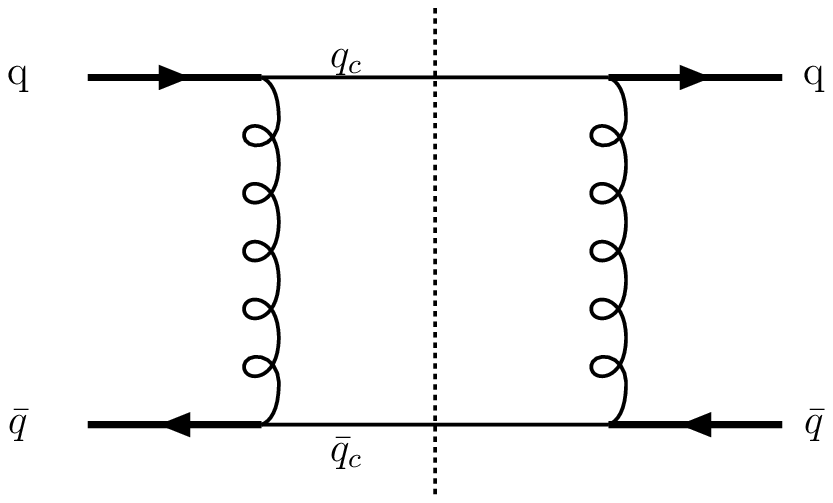} 
\end{center}
\caption{ Modified diagrammatic for $q-\bar q$ scattering in L-N model.
\label{f3}}
\end{figure}

As both $q$ and $q_c$ are color triplet with the same flavor, thus $c$ is a  color 
octet flavorless object. There is no prior reason to suppose $c(\bar c)$ is a 
fundamental constituent. It is just a nonperturbative gluer, but its concrete form 
has been determined phenomenologically by $L-N$ in \cite{ln}. Following L-N model which identifies 
IP exchange as nonperturbative two-gluon exchange. We further think the IP is a 
colorless and flavorless object which composed from a conjugate pair of such color 
octet nonperturbative gluer, $c\bar c$.

Before discussing other relevant problems, let us observe an interesting property for 
Fig.3, i.e. when its cutting current-quark lines, $q_c$(or $\bar q_c$), are replaced 
by $q+\bar c$ (or $\bar q +c$) lines (a crossing process of MNSIR!) then Fig.3 going 
into Fig.4 and we further draw it as Fig.5

\begin{figure}[htb]
\begin{minipage}[t]{.45\linewidth}
\begin{center}
%\vskip -5mm
\epsfxsize=12pc
\epsfbox{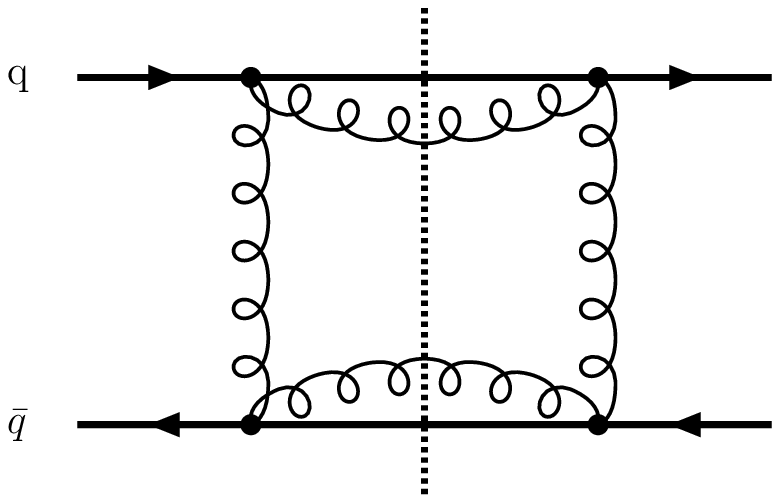}
\end{center}
\caption{ The equivalent diagram of Fig.3 after a crossing process of MNSIR.
\label{f4}}
\end{minipage}%
\hspace{0.5in}
\begin{minipage}[t]{.45\linewidth}
\begin{center}
\epsfxsize=12pc
\epsfbox{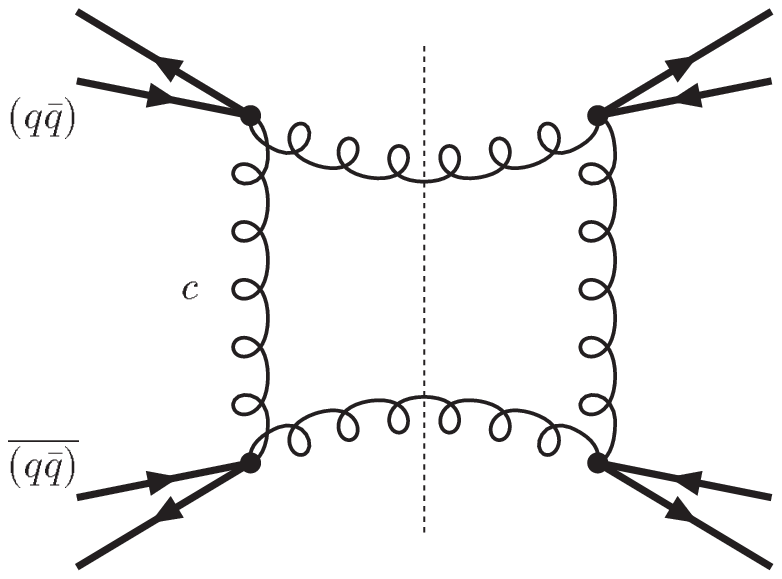}
\end{center}
\caption{ The same topology as fig.4.
\label{f5}}
\end{minipage}
\end{figure}

We would emphasize such operation is possible just because these $q_c$ (or $\bar q_c$) 
lines are on mass shell, thus there is no interaction between the splitted $q$ and 
$\bar c$ (or $\bar q$ and $c$). The only effect of such splitting is sharing the 
energy-momentum of original cut  $q_c$ $(\bar q_c)$ lines.

Now let us consider what would happens when Fig.4 were embedded into any high energy 
$h-h'$ scatting. If the additive quark rule in total cross section is valid exactly 
(in fact the uncertainty of this rule is about 10$\%$), then we just need to embed Fig.4 
into Fig.1, thus the $q$ (or $\bar q$) does not couple to exchanged gluons and plays 
only as spectra-particles to share the energy-momentum of the whole system and provide 
the number of members in a hadron for additive quark rule, but have nothing to do 
with the major dynamics reaction  in  scattering.

In discussing the crossing channel physical problems relevant with above observation, 
the most convenient way is to take  parallel comparison between Regge pole scheme 
with our conjecture or demonstration about IP, constituent gluon ($g$) and glueball 
states ($G's$).See Fig.c

\begin{figure}[t]
%\figurebox{20pc}{15pc}{} % to have a box alone
\begin{center}
\epsfxsize=13cm 
\epsfbox{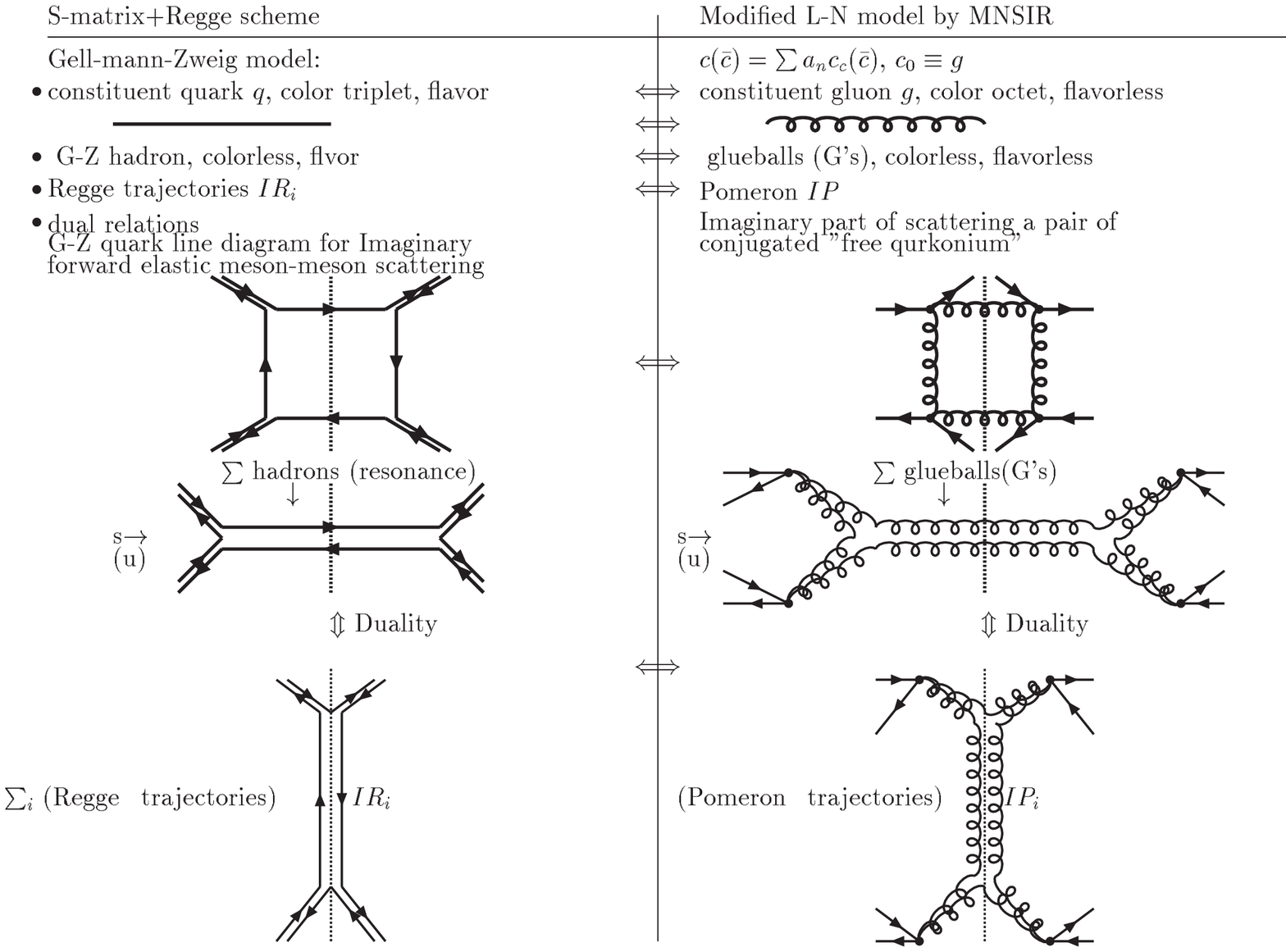} 

Fig.c
\end{center}
\end{figure}
In Regge pole theory there is an average duality, which means the integral of the 
imaginary part of all hadrons and resonances (including continuum from $q$ ($\bar q$)) 
in $s$-(or $u-$) channel of a scattering amplitude is equal to that of all Regge pole 
trajectories contributions in $t$-channel. We think the average duality between 
glueballs (including continuum from $g$) in $s$-(or $u$-) channel with IP in 
$t$-channel may also be valid as shown in above comparison.

Finally, we  briefly discuss about the mechanism of strong-production of glueball. 
We think it could only be produced in high energy strong-soft processes, since  the 
production of glueballs are certainly produced from non-pertubative interaction and the 
constituent gluons are abundant in high energy strong-soft process. Glueballs are 
produced in such process via the fusion of colliding constituent gluon pair. The 
double diffractive scattering process $p+p(\bar p)\rightarrow p+p(\bar p)+G$ would 
have a very rich yield for $G$ production\cite{pxb}  \cite{pdh}.

\end{document}